\newif\ifarxiv\arxivtrue
\ifarxiv\else
  \PassOptionsToClass{linenumbers}{revtex4-2}
\fi
\documentclass[%
 reprint,
 superscriptaddress,
 amsmath,amssymb,
 aps, physrev
]{revtex4-2}

\usepackage{graphicx, mathtools}
\usepackage{dcolumn}
\usepackage{bm}
\usepackage{dsfont}
\usepackage{color,xcolor}
\usepackage{import}

\usepackage{transparent}
\usepackage[nameinlink]{cleveref}
\makeatletter\let\H@refstepcounter\refstepcounter\makeatother
\crefname{figure}{Fig.}{Figs.}
\Crefname{figure}{Figure}{Figures}
\crefname{equation}{Eq.}{Eqs.}
\Crefname{equation}{Equation}{Equations}
\crefname{section}{Sec.}{Secs.}
\Crefname{section}{Section}{Sections}
\crefname{subsection}{Sec.}{Secs.}
\Crefname{subsection}{Section}{Sections}
\crefname{table}{Table}{Tables}
\Crefname{table}{Table}{Tables}
\crefname{appendix}{Appendix}{Appendices}
\Crefname{appendix}{Appendix}{Appendices}
\ifarxiv
  \newcommand{\fj}[1]{}
  \newcommand{\fd}[1]{}
  \newcommand{\sgj}[1]{}
  \newcommand{\rp}[1]{}
\else
  \newcommand{\fj}[1]{{\color{olive}{[fj: #1]}}}
  \newcommand{\fd}[1]{{\color{red}{[fd: #1]}}}
  \newcommand{\sgj}[1]{{\color{blue}{[sgj: #1]}}}
  \newcommand{\rp}[1]{{\color{yellow}{[rp: #1]}}}
\fi

\newcommand{\loss}{\mathcal{L}}

\makeatletter
\DeclareRobustCommand\equalcontrib{\global\let\equalcontrib@mark\equalcontrib@dagger}
\def\equalcontrib@dagger{{\comma@space}\frontmatter@footnotemark{2}}
\let\equalcontrib@mark\@empty
\def\doauthor#1#2#3{%
  \ignorespaces#1\unskip\@listcomma
  \begingroup
   #3%
  \@if@empty{#2}{\endgroup{}{}}{\endgroup{\comma@space}{}\frontmatter@footnote{#2}}%
  \equalcontrib@mark\global\let\equalcontrib@mark\@empty
  \space \@listand
}%
\makeatother

\begin{document}

\preprint{APS/123-QED}

\title{\textbf{Optimization of Active Reconfigurable Surfaces for Spatiotemporal MU-MIMO Wave Routing in Disordered Media: Application to the Cocktail-Party Problem} 
}%

\author{Frederic J. N. Jorgensen\equalcontrib}
\email{Contact author: fjorgen@mit.edu}
\affiliation{Department of Mathematics, Massachusetts Institute of Technology, Cambridge, Massachusetts 02139, USA}
\author{Florent Dorlot}
\thanks{These authors contributed equally to this work.}
\author{Constant Bourdeloux}
\author{Mathias Fink}
\author{Fabrice Lemoult}
\affiliation{Institut Langevin, ESPCI Paris, PSL University, CNRS, Paris, France}
\author{Steven G. Johnson}
\affiliation{Department of Mathematics, Massachusetts Institute of Technology, Cambridge, Massachusetts 02139, USA}
\author{Raphaël Pestourie}
\affiliation{School of Computational Science and Engineering, Georgia Institute of Technology, Atlanta, Georgia 30332, USA}


\date{\today}

\begin{abstract}
We design an active reconfigurable intelligent surface (RIS) for acoustic multi-user multiple-input multiple-output (MU-MIMO) communication by solving a large-scale convex optimization problem, so that each emitter's broadband signal focuses in space and time at its intended receivers while crosstalk is suppressed. In acoustic experiments, we show that our approach outperforms existing time-reversal methods by 5.1--7.6\,dB in crosstalk suppression, while achieving better temporal focusing.
\end{abstract}

\maketitle

\begin{figure}[b]
\includegraphics[width=\columnwidth]{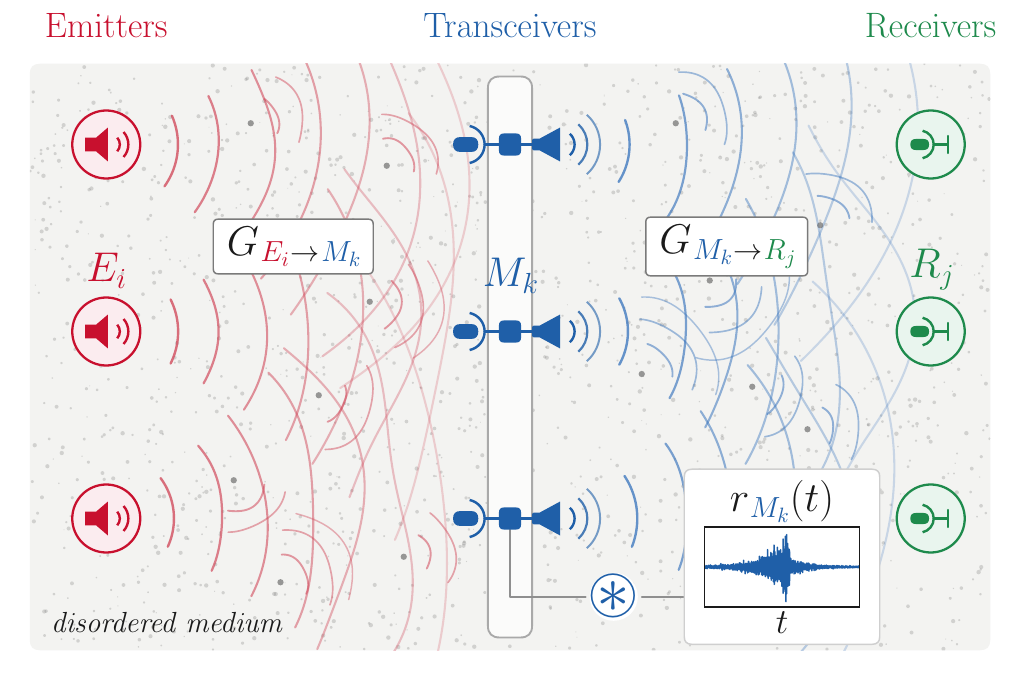}
\caption{\label{fig:setup}
Schematic MU-MIMO problem. Emitters $E_i$ reach receivers $R_j$ through a disordered medium and an active surface of transceivers~$M_k$. Each $M_k$ re-emits the wave it records, convolved with its kernel $r_{M_k}(t)$. Green's functions $G_{E_i \rightarrow M_k}$ and $G_{M_k \rightarrow R_j}$ describe propagation to and from the surface.}
\end{figure}

We consider the problem of designing reconfigurable intelligent surfaces (RIS) for broadband multi-user multiple-input multiple-output (MU-MIMO) communication and show that designs obtained by large-scale convex optimization achieve lower crosstalk and better temporal focusing than previous time-reversal designs. We developed this method with particular emphasis on acoustic applications, a setting commonly referred to as the \textit{cocktail-party problem}~\cite{cherry1953some, arons1992review, bourdeloux2024solution, Dorlot2025, pestourie2024towards}. In this setting, emitters and receivers sit in a disordered medium and communicate mainly via an RIS, whose designable units record the incoming field and re-emit it after filtering (\cref{fig:setup}).  Given measurements of the Green's function (impulse response) between emitters or receivers and the RIS, the goal of the cocktail-party problem is to design the RIS filters to maximize the undistorted transmission between desired pairs and minimize the crosstalk to other receivers.

Previous designs of these filters relied on time reversal (TR)~\cite{bourdeloux2024solution, Dorlot2025}, in which the recorded Green's function is time-reversed and used as a matched filter~\cite{fink1999time, fink2002time, stojanovic2005retrofocusing}.  If one could record the \emph{entire} wavefield on a surface enclosing a source placed at the receiver and time-reverse it, that would focus waves on the receiver up to the diffraction limit~\cite{fink2002time, derosny2002overcoming}, but such a closed time-reversal cavity is not possible here because of the limited RIS aperture. In practice, therefore, one approximates the TR signal using partial data (a finite aperture) at the RIS locations. TR has two other limitations.  Most importantly, a TR RIS maximizes the signal at the intended receivers but does not necessarily eliminate crosstalk to other receivers. Second, to target multiple receivers simultaneously, simply superposing multiple TR designs introduces additional distortion due to this crosstalk.

Here, we instead design the filters of all transceivers jointly by formulating crosstalk-suppressing MU-MIMO design as a weighted least-squares problem over causal finite impulse response (FIR) filters and solving the resulting convex quadratic optimization problem with a matrix-free conjugate-gradient (CG) method whose computational cost per iteration (milliseconds on a CPU) does not grow with the number of receivers. 

\begin{figure*}[t]
    \centering
    \includegraphics[width=\textwidth]{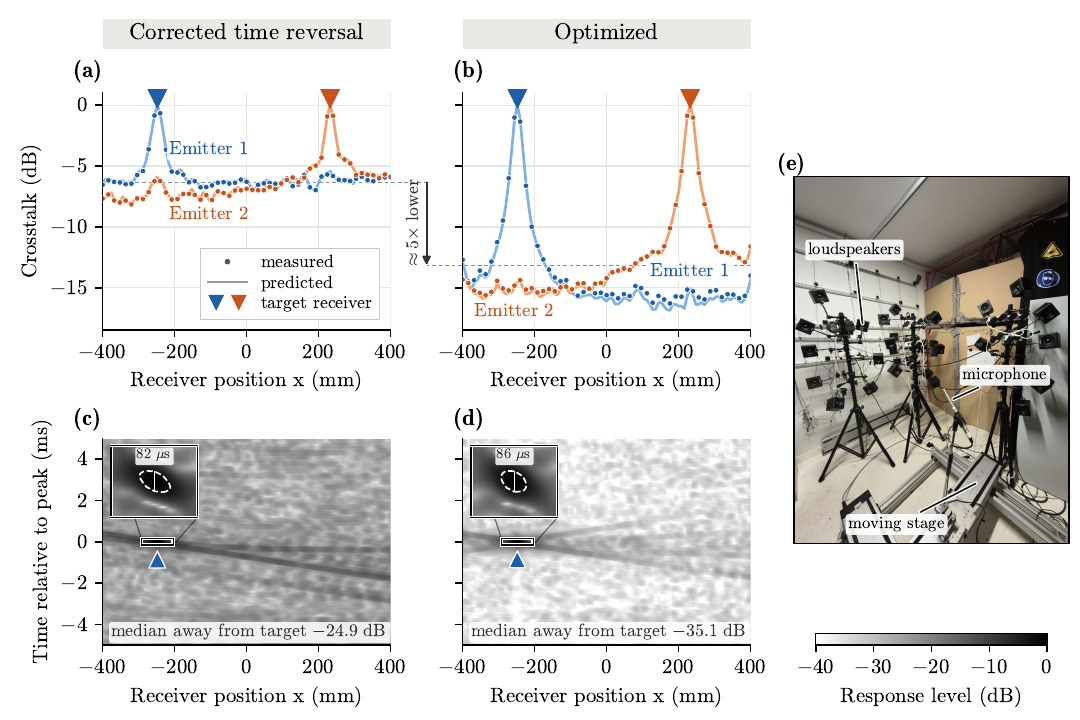}
    \caption{\label{fig:experiment_result}
    Focusing each emitter on one receiver, with corrected TR (left) and optimized filters (right) from \cref{eq:lstq_objective} with $c = 0.1$, $N_M = 10{,}000$ and 200 conjugate-gradient (CG) iterations. Triangles mark the targets. (a), (b) Crosstalk along the receiver line, both predicted (lines) and measured (dots). Dashed lines give the mean crosstalk over all receivers except the two on each side of a target, and the arrow its reduction. (c), (d) Emitter~1 at all receivers around its arrival, after pulse compression, as a 0.25\,ms moving RMS relative to the focus, against time relative to the peak at the target. Insets: envelope at the focus with the FWHM of its power. (e) Receiver side of the setup.}
\end{figure*}

We demonstrate our approach on the reverberant broadband acoustic experiment (120\,Hz to 12\,kHz) of Dorlot et~al.~\cite{Dorlot2025}, with 2~emitters, 101~receiver positions and an RIS of 28~transceivers. Compared with TR, the optimized filters reduce crosstalk by 5.1--7.6\,dB (a factor of 3--6), and lower the residual temporal energy around the arrival by about 12\,dB using filters of the same length (\cref{fig:experiment_result}). A model built from the measured Green's functions predicts the measured mean crosstalk to within 0.3\,dB, and a complete design takes 0.4\,s (for 200 CG iterations) on a laptop CPU (Apple M5 Pro).

\textit{Problem setting.} The configuration under consideration is illustrated in \cref{fig:setup} and is similar to the setups recently considered by Bourdeloux et~al.~\cite{bourdeloux2024solution} and Dorlot et~al.~\cite{Dorlot2025}. It consists of $n_E$ emitters $\mathcal{E} = \{E_i\}_{i=1}^{n_E}$ and $n_R$ receivers $\mathcal{R}  = \{R_j\}_{j=1}^{n_R}$, each at a fixed position.  Our objective is to simultaneously establish broadband communication channels between selected emitter--receiver pairs $\mathcal{P}\subseteq \mathcal{E} \times \mathcal{R}$, while suppressing communication between all other pairs $\mathcal{P}^C$.  A key assumption underlying our modeling is that wave propagation between an emitter $E_i$ and a receiver $R_j$ is predominantly mediated by an active RIS consisting of transceivers $\mathcal{M} = \{M_k\}_{k=1}^{n_M}$ that each act as a linear, causal, and time-invariant filter. In practice, this is achieved by a sufficiently high gain of the transceivers compared to other propagation paths. 
Accordingly, each transceiver is described by its impulse response $r_{M_k}(t)$, which we call its kernel. These kernels are our design parameters. 
For linear wave propagation, the system's total response function $ h_{E_i\rightarrow R_j} $ from emitter $E_i$ to receiver $R_j$ can be described in terms of Green's functions $G_{E_i \rightarrow M_k}(t)$ and $G_{M_k\rightarrow R_j}(t)$ between the emitters, transceivers, and receivers. It takes the form~\cite{Dorlot2025} 
\begin{equation}
\label{eq:response_function}
    h_{E_i\rightarrow R_j} \coloneqq \sum\limits_{k } G_{E_i \rightarrow M_k \rightarrow R_j} \circledast r_{M_k},
\end{equation}
where $\circledast$ is the temporal convolution and $G_{E_i \rightarrow M_k \rightarrow R_j} \coloneqq G_{E_i \rightarrow M_k} \circledast G_{M_k\rightarrow R_j}$ is the composite Green's function from emitter $E_i$ through transceiver $M_k$ to receiver $R_j$. 

\textit{Related work.} At first glance, wavefront shaping in \textit{disordered} or strongly scattering media~\cite{fink2002time, lerosey2004time, lerosey2007focusing, vellekoop2007focusing, vellekoop2008universal, mosk2012controlling, rotter2017light, frazier2020wavefront} may appear more challenging than in free space~\cite{braun1905strassburger, yagi1928beam, capon1969high} due to the complex multiple scattering of waves. The opposite is true: disorder typically facilitates control because the Green's function of a disordered medium supports a large number of independent transmission channels~\cite{derode2001random, blomgren2002super, lemoult2009manipulating, rotter2017light}. Such media offer \textit{greater} flexibility for wave manipulation than free space, whose Green's function is highly correlated and supports only a few independent channels~\cite{greengard1987fast}. An RIS is a recent approach to exploit these many channels~\cite{bourdeloux2024solution, zhang2024optimizing, Dorlot2025}. 

For a single emitter $E$ and a single receiver $R$, Bourdeloux et~al.~\cite{bourdeloux2024solution} and Dorlot et~al.~\cite{Dorlot2025} propose solving this single-input single-output (SISO) case of the problem posed above with the time-reversal filters $r_{M_k} = G^{\mathrm{TR}}_{E\rightarrow M_k}\circledast G^{\mathrm{TR}}_{M_k\rightarrow R}$, where $G^{\mathrm{TR}}_{E\rightarrow M_k}(t)\coloneqq G_{E\rightarrow M_k}(T_E-t)$ and $G^{\mathrm{TR}}_{M_k\rightarrow R}(t)\coloneqq G_{M_k\rightarrow R}(T_R-t)$, with global time shifts $T_E$ and $T_R$. The Green's functions are truncated to finite time windows, since they are measured and sampled over a finite duration. Choosing $T_E$ and $T_R$ as the lengths of these windows makes the time-reversed kernels vanish for $t<0$, so they are causal by construction. 
Time-reversal methods exploit the time-reversal symmetry of the wave equation (only approximate in dissipative media): a wavefield recorded at the transceivers and re-emitted time-reversed retraces its multiple-scattering paths and refocuses at its source~\cite{fink2002time, fink1999time}. Combining both time-reversed sides, $r_{M_k}$ thus routes a signal from $E$ through all transceivers to $R$, where all paths add up coherently. TR is also a matched filter, which maximizes the output signal-to-noise ratio for additive white noise~\cite{turin1960introduction}.  

For multiple pairs $(E_i,R_j)\in\mathcal{P}$, the single-pair kernels are superposed~\cite{bourdeloux2024solution,Dorlot2025},
\begin{equation}
\label{eq:tr_coeffs}
r_{M_k} = \sum\limits_{(E_i,R_j)\in\mathcal{P}} G^{\mathrm{TR}}_{E_i\rightarrow M_k\rightarrow R_j},
\end{equation}
where $G^{\mathrm{TR}}_{E_i\rightarrow M_k\rightarrow R_j}\coloneqq G^{\mathrm{TR}}_{E_i\rightarrow M_k}\circledast G^{\mathrm{TR}}_{M_k\rightarrow R_j}$ is the composite time-reversed Green's function of the pair $(E_i,R_j)$. This enables multipoint spatiotemporal focusing, as demonstrated underwater~\cite{bourdeloux2024solution} and, on an active RIS in air, by Dorlot et~al.~\cite{Dorlot2025}, whose experimental platform we use here. 
To improve signal fidelity despite the limited bandwidth of the microphones, loudspeakers, and room, Dorlot et~al.\ additionally equalize the kernels in frequency~\cite{Dorlot2025}: the spectrum of each kernel in \cref{eq:tr_coeffs} is multiplied by a real gain $H(\omega)$, obtained from a preliminary focusing test such that the smoothed spectrum of the focused signal matches that of the probe chirp. We refer to these equalized kernels as \emph{corrected time reversal} (corrected TR) and use them as the experimental baseline. 

TR remains limited by crosstalk, which also degrades superposition, and by signal distortion, both worsened by the finite aperture, bandwidth, and dissipation. We address these limitations by optimizing the filters of all transceivers jointly in the time domain. Since each transceiver filters the unknown superposition of all emitters in real time, a single design must route \textit{all} emitters at once with causal filters of fixed length, which couples all frequencies. This sets our problem apart from related work. Inverse filtering focuses better than TR in space and time by inverting the Green's function between a \textit{source} array and chosen control points one frequency at a time~\cite{tanter2000time, tanter2001optimal, yon2003sound}, but such per-frequency inversion gives filters that are neither causal nor of fixed length, and are delicate to regularize. Iterative TR approximates the inverse filter more robustly~\cite{montaldo2004real, montaldo2004telecommunication}. Our earlier per-frequency design~\cite{pestourie2024towards} performed poorly in experiments once fitted with FIR filters. Another system is a passive RIS, which is limited to $\pm 1$ reflection coefficients (leading to challenging binary-optimization problems)~\cite{zhang2024optimizing}. Perhaps the closest formulation to ours was a least-squares time-domain FIR design for loudspeaker arrays in personal audio~\cite{simongalvez2015time} (targeting low distortion in a ``bright zone'' with low crosstalk in a ``dark zone''), but in this case there was only a single emitter (wired directly to the transceivers), and the system was small enough to allow direct (non-scalable) matrix solvers (2--8 loudspeakers and 16--512 filter taps). Moreover, the performance of those filters was only computed from measured responses, while we measure ours experimentally in a reverberant medium.

\textit{Method.} We present a method to design the transceiver kernels $r_{M_k}$ that route each emitter to its selected receivers while suppressing crosstalk to all others. We formulate this problem as a weighted least-squares problem and solve it efficiently with a conjugate-gradient solver.

In terms of the system's overall transfer function $\hat{h}_{E_i \rightarrow R_j}(\omega)$, the Fourier transform of the time-domain response $h_{E_i \rightarrow R_j}(t)$ of \cref{eq:response_function}, we require each routed pair to match a prescribed flat target transmission function, $\hat{h}_{E_i \rightarrow R_j}(\omega)\approx \hat{h}^{\mathrm{ref}}_{E_i\rightarrow R_j}(\omega)$ for all $(E_i, R_j) \in \mathcal{P}$. This preserves the waveform of the input signal, as needed when the receiver (such as a human listener) cannot rely on signal post-processing. Telecommunication systems with extensive post-processing instead typically maximize the Shannon channel capacity~\cite{shannon1948mathematical}. Suppressing crosstalk between all other pairs means that $\hat{h}_{E_i \rightarrow R_j}(\omega) \approx 0$ for all $(E_i, R_j) \notin \mathcal{P}$. 

This motivates solving the MU-MIMO problem by minimizing the following objective (loss) functional $\loss$ over the kernels $r_{M_k}$:
\begin{align*}
\loss\left(\left\{r_{M_k}\right\}_{k = 1}^{n_M}\right) & \coloneqq \sum\limits_{(E_i,R_j) \in \mathcal{P}}\left\|h_{E_i \rightarrow R_j}- h^{\mathrm{ref}}_{E_i\rightarrow R_j} \right\|_2^2 \\
&+c \sum\limits_{(E_i,R_j) \not \in \mathcal{P}}\left\|h_{E_i \rightarrow R_j} \right\|_2^2
\end{align*}
for a tradeoff parameter $c\geq 0$ and the $L_2$ norm $\|\cdot\|_2$. 
By Parseval's theorem, this objective may be evaluated in the time or the frequency domain. It is convenient  to rewrite $\loss$ in the more compact form 
\begin{equation}
\label{eq:lstq_objective}
    \loss\left(\left\{r_{M_k}\right\}_{k = 1}^{n_M}\right) = \sum\limits_{i,j}w_{ij}\left\|h_{E_i \rightarrow R_j}- h^{\mathrm{ref}}_{E_i\rightarrow R_j} \right\|_2^2 
\end{equation}
with $w_{ij}\coloneqq c + (1-c)\cdot \mathds{1}_{\mathcal{P}}\left(E_i,R_j\right)$ for the indicator function $\mathds{1}_{\mathcal{P}}$ and  $ h^{\mathrm{ref}}_{E_i\rightarrow R_j} = 0$  for all $(E_i, R_j) \not\in\mathcal{P}$. Because each response $h_{E_i\rightarrow R_j}$ depends linearly on the kernels  $r_{M_k}$, $\loss$ is a convex quadratic functional, and convexity makes finding its global optimum tractable.  

\textit{Discretization.} In practice, each transceiver implements an FIR filter with $N_M$ real coefficients $r_{kn}$ and clock period $\Delta t$,
\begin{equation}
\label{eq:fir}
r_{M_k}(t) = \sum\limits_{n = 0}^{N_M-1} r_{kn}\,\delta(t - n\Delta t),
\end{equation}
and the Green's functions are known only as the sampled sequences $\{G_{E_i\rightarrow M_k}(n\Delta t)\}_{n=0}^{N_E-1}$ and $\{G_{M_k\rightarrow R_j}(n\Delta t)\}_{n=0}^{N_R-1}$ of lengths $N_E$ and $N_R$. With $\langle\cdot,\cdot\rangle$ the $L_2$ inner product, the first-order optimality condition of \cref{eq:lstq_objective} requires
\begin{equation}
\label{eq:variational_equation}
\sum\limits_{i,j} w_{ij}\big\langle h_{E_i\rightarrow R_j} - h^{\mathrm{ref}}_{E_i\rightarrow R_j},\, G_{E_i\rightarrow M_k\rightarrow R_j}\circledast\phi\big\rangle = 0
\end{equation}
for all test functions $\phi$ and transceivers $M_k$. We discretize \cref{eq:variational_equation} with a Galerkin method, which here coincides with a Nyström method, using the FIR basis of \cref{eq:fir} as trial functions and the impulses $\phi_n(t) = \delta(t - n\Delta t)$ as test functions, and evaluate the integrals by quadrature on the sampling grid, so that $\circledast$ becomes discrete convolution. This yields the linear system of normal equations
\begin{align}
U r &= b, \label{eq:variation_equation_finite}\\
U_{kn,k'n'} &= \sum\limits_{i,j} w_{ij} \big\langle G_{E_i \rightarrow M_k \rightarrow R_j} \circledast \phi_n, \nonumber\\
&\hphantom{{}= \sum\limits_{i,j} w_{ij} \big\langle}\, G_{E_i \rightarrow M_{k'} \rightarrow R_j} \circledast \phi_{n'}\big\rangle, \label{eq:U_definition}\\
b_{kn} &= \sum\limits_{i,j} w_{ij} \big\langle h^{\mathrm{ref}}_{E_i\rightarrow R_j},\, G_{E_i \rightarrow M_k \rightarrow R_j} \circledast \phi_n\big\rangle. \label{eq:b_definition}
\end{align}

\textit{Computational details.} Forming $U$ explicitly is expensive: for the experiment below, with $n_M = 28$ and $N_M = 10{,}000$, it has dimension $280{,}000 \times 280{,}000$ and would require about 630\,GB of memory in double precision. Since $U$ is positive semidefinite, we instead apply CG to the normal equations (CGNR)~\cite{saad2003iterative}, which only requires evaluating products $Uv$ for $v \in \mathbb{R}^{n_M\times N_M}$. These can be written as
\begin{align}
(Uv)_{kn} &= \sum\limits_{k'} \big(K_{k'k} \circledast v_{k'\cdot}\big)(n), \label{eq:efficient_uv_computation}\\
K_{k'k} &\coloneqq \sum\limits_{i,j} w_{ij}\, G^\dagger_{E_i\rightarrow M_k\rightarrow R_j} \circledast G_{E_i\rightarrow M_{k'}\rightarrow R_j}, \nonumber
\end{align}
where $v_{k'\cdot} \coloneqq (v_{k'n})_{n=0}^{N_M-1}$. Thus $U$ is block Toeplitz, $U_{kn,k'n'} = K_{k'k}(n-n')$, and each product is a sum of convolutions. Here $f^\dagger(t) \coloneqq \overline{f(-t)}$ defines the adjoint of convolution, $(f\circledast)^\dagger = f^\dagger\circledast$. Evaluating all convolutions with a fast Fourier transform (FFT) on a grid of length $N_{\mathrm{tot}} \coloneqq N_E + N_R + N_M - 2$, the length of each response $h_{E_i\rightarrow R_j}$, one product takes $2n_M$ FFTs and one $n_M\times n_M$ matrix--vector product per frequency, i.e.\ $\mathcal{O}(n_M N_{\mathrm{tot}}\log N_{\mathrm{tot}} + n_M^2 N_{\mathrm{tot}})$ operations. The FFTs of all $K_{k'k}$ are computed only once. Because $w_{ij} = c + (1-c)\,\mathds{1}_{\mathcal{P}}(E_i,R_j)$, their sum over all pairs factorizes at each frequency into $c$ times the elementwise product of an emitter-side and a receiver-side $n_M\times n_M$ Gram matrix, plus a sum over the routed pairs. This precomputation, including $b$, costs $\mathcal{O}\big((n_E + n_R + |\mathcal{P}|)\, n_M^2 N_{\mathrm{tot}}\big)$ plus the FFTs of all Green's functions, so it grows with $n_E + n_R$ rather than with the number $n_E n_R$ of emitter--receiver pairs, and the cost of a CG iteration does not depend on $n_E$ or $n_R$ at all. The total is orders of magnitude below the $\mathcal{O}\big((n_M N_M)^3\big)$ cost of a dense direct solve.

We note an interesting connection between the iterative optimizer and TR: for a target pulse $h^{\mathrm{ref}}_{E_i\rightarrow R_j}(t)=\delta(t-T_0)$ at delay $T_0$, the right-hand side of \cref{eq:variation_equation_finite} becomes $b_{kn} = \sum\limits_{(E_i,R_j)\in\mathcal{P}} G_{E_i\rightarrow M_k\rightarrow R_j}(T_0-n\Delta t)$, which is the superposed time-reversal kernel of \cref{eq:tr_coeffs} when $T_0 =T_E+T_R$. Started from $r=0$, CG therefore returns TR, scaled to minimize $\loss$, as its first iterate. Each further iteration applies $U$, which plays the current kernels through the medium to all receivers and time-reverses the weighted responses back to the transceivers. The $m$-th CG iterate minimizes $\loss$ over the Krylov subspace spanned by $\{ b, Ub, \dots, U^{m-1}b \}$~\cite{saad2003iterative}, which also contains the $m$-th iterate of ``iterative TR''~\cite{montaldo2004real} (gradient descent on $\loss$ with a fixed step, also started from $r=0$). After any number of iterations, CG thus reaches a loss no higher than iterative TR, and both converge to the minimizer of \cref{eq:lstq_objective}.

\textit{Conditioning and regularization.} $U$ is ill-conditioned, with $\kappa(U)\geq3\times10^{16}$, which requires regularization. One option is a Tikhonov penalty $\lambda\|r\|_2^2$~\cite{simongalvez2015time,hansen1998rank}. We instead stop CG early, which, starting from $r=0$, has a comparable regularizing effect~\cite{hansen1998rank} and performed similarly to Tikhonov in our experiments. Unlike Tikhonov, early stopping requires no converged solve for each $\lambda$, and it keeps the connection above to TR. We calibrate the iteration count on the measured data. After 200 iterations, the Lanczos matrix of CG has a condition number of only $2.8\times10^4$. 

The discretized objective of \cref{eq:lstq_objective} is $\|Ar - d\|_2^2$, where $Ar$ and $d$ stack the weighted responses $\sqrt{w_{ij}}\,h_{E_i\rightarrow R_j}$ and targets $\sqrt{w_{ij}}\,h^{\mathrm{ref}}_{E_i\rightarrow R_j}$. Forming the normal equations squares its condition number~\cite{saad2003iterative}, which amplifies round-off errors but does not change how measurement errors propagate, because both formulations share the same minimizer. An iterative least-squares solver such as LSQR~\cite{paige1982lsqr} avoids the squaring by minimizing $\|Ar - d\|_2^2$ directly, with the same iterates as CG in exact arithmetic. Each of its iterations, however, costs $\mathcal{O}(n_E n_R n_M N_{\mathrm{tot}} + n_M N_{\mathrm{tot}}\log N_{\mathrm{tot}})$, compared with $\mathcal{O}(n_M^2 N_{\mathrm{tot}} + n_M N_{\mathrm{tot}}\log N_{\mathrm{tot}})$ for $Uv$, so the normal equations are cheaper in computation and memory whenever $n_E n_R \gg n_M$, as in our experiments. On the measured data of \cref{fig:experiment_result}, the filters of both solvers differed only by accumulated rounding errors (relative difference $2\times10^{-12}$ up to 60 iterations and $4\times10^{-3}$ after 200 iterations), and their predicted crosstalk agreed to within 0.01\,dB, but our normal-equation implementation was 17 times faster (0.40\,s versus 6.9\,s for 200 iterations).
\begin{figure*}[t]
\centering
\includegraphics[width=\textwidth]{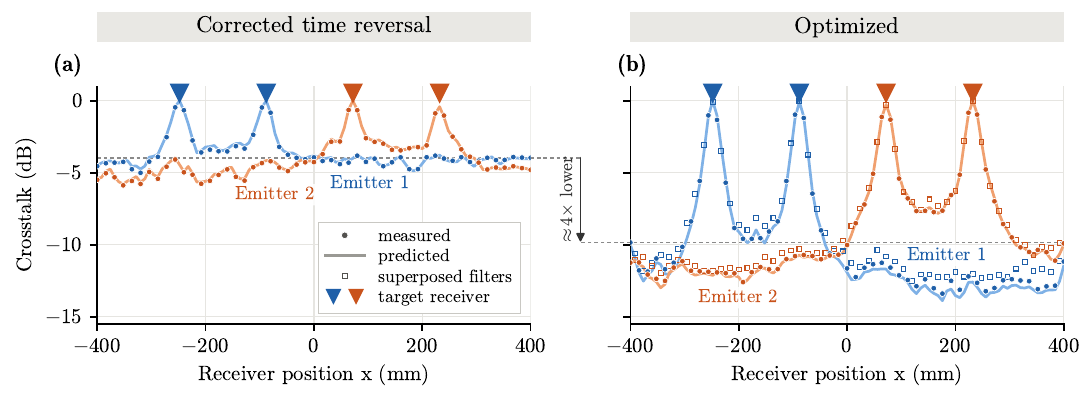}
\caption{\label{fig:experiment_2_pairs}
Focusing each emitter on two receivers. (a) Corrected TR. (b) Optimized filters with $c = 0.1$, $N_M = 10{,}000$ and 200 CG iterations. As in \cref{fig:experiment_result}, dots are measured, lines are predicted, and triangles mark each emitter's target receivers. Squares in (b) are the measured crosstalk of the superposed filters, i.e.\ the sum of the two filters optimized separately for one target each.}
\end{figure*}

\textit{Results.} To demonstrate our method, we use the reverberant acoustic cocktail-party experimental setup of Dorlot et~al.~\cite{Dorlot2025}, on which they demonstrated time-reversal routing, here with new loudspeakers and a single audio interface (Antelope Orion 32+). It consists of two rooms, one with $n_E = 2$ emitters and the other with $n_R = 101$ receiver positions spaced 8\,mm apart along a line, acoustically coupled through an RIS of $n_M = 28$ transceivers, each a microphone on the emitter side and a loudspeaker on the receiver side implementing the causal response $r_{M_k}(t)$. We measure the Green's functions $G_{E_i \rightarrow M_k}$ and $G_{M_k \rightarrow R_j}$ by transmitting a 1\,s linear chirp $p(t)$ from 120\,Hz to 12\,kHz, sampled at 44.1\,kHz, and reconstruct them with a Wiener filter to length $N_E = N_R = 7680$. From these Green's functions we compute the filters with $c = 0.1$, $N_M = 10{,}000$ and 200 CG iterations, values chosen after testing others on the setup. The target response $h^{\mathrm{ref}}$ is a Tukey-windowed~\cite{prabhu2014window} band-pass pulse between 120\,Hz and 12\,kHz that peaks at $T_0 = 136$\,ms. This delay sets the latency, 91\,ms less than for corrected TR, whose kernels focus at their last tap. Truncating the Green's functions used for TR gives it the same latency at a predicted cost of less than 0.1\,dB in crosstalk. 
In the first experiment, each emitter $E_i$ is assigned a single target receiver $R_j$. To assess spatial focusing, we transmit the probe chirp $p(t)$ through each emitter $E_i$ ($i = 1, 2$) and record the responses at all receiver positions. The crosstalk levels for the two methods are shown in \cref{fig:experiment_result}(a) and (b), where crosstalk is the energy of the recorded signal, cross-correlated with the probe chirp, within $\pm 11.3$\,ms of the arrival at the target receiver, normalized to the target receiver. Compared with the corrected TR baseline [\cref{fig:experiment_result}(a)], the optimized filters [\cref{fig:experiment_result}(b)] lower the mean crosstalk by 6.0--7.6\,dB. The crosstalk predicted by convolving the filters with the measured Green's functions (lines) agrees with the measurement to within 0.3\,dB in the mean for both methods, with outliers of up to 0.8\,dB. To assess temporal focusing, \cref{fig:experiment_result}(c) and (d) show the response of emitter~1 at all receivers around its arrival at the target, after pulse compression. Both designs produce a similarly short focal spot. At the target, the full width at half maximum (FWHM) of its power envelope is 82\,$\mu$s for TR and 86\,$\mu$s for the optimized filters, close to the 80\,$\mu$s expected by a 12\,kHz passband. Around the focus, however, TR leaves a stronger wavefront sweeping across the receivers and a higher background signal, with a median level away from the target of $-24.9$\,dB versus $-35.1$\,dB for the optimized filters (lower left of the maps). TR favors the strongest paths, the direct sound of the transceivers reaching the receivers from one side, whereas the optimized filters rely more on reflections from the opposite side, visible in \cref{fig:experiment_result}(d) as a second wavefront crossing the focus. At the target itself, the median level between 2 and 25\,ms before and after the arrival lies 29--30\,dB below the peak for TR and 40--43\,dB below for the optimized design (ranges over both emitters, before and after), i.e.\ roughly 12\,dB lower. Emitter~2 (not shown) behaves similarly, with FWHMs of 81 and 86\,$\mu$s and a median level away from the target 8.5\,dB lower with the optimized filters. 

To show that our design method extends to more general configurations, we perform an additional experiment in which each emitter targets 
\emph{two} receivers, following the same experimental and optimization procedure as before.  The results, shown in \cref{fig:experiment_2_pairs},  confirm that the optimized filters again focus better in space, lowering the mean crosstalk, taken relative to the weaker target, by 5.1--6.6\,dB compared with TR. 

An alternative way to target multiple receivers is to superpose the filters optimized separately for each target~\cite{bourdeloux2024solution, Dorlot2025}. Given low crosstalk, this is nearly equivalent to the joint design. Split $\mathcal{P}$ into disjoint subsets $\mathcal{P}_m$ and let $r^{(m)}$ minimize \cref{eq:lstq_objective} for the pairs $\mathcal{P} = \mathcal{P}_m$. Then $\sum_m r^{(m)}$ exactly minimizes \cref{eq:lstq_objective} with the target $h^{\mathrm{ref}}_{E_i\rightarrow R_j}$ of each pair $(E_i,R_j)\in\mathcal{P}$ replaced by $h^{\mathrm{ref}}_{E_i\rightarrow R_j}+(1-c)\sum_{m'\neq m}h^{(m')}_{E_i\rightarrow R_j}$, where $\mathcal{P}_m$ is the subset containing $(E_i,R_j)$ and $h^{(m')}_{E_i\rightarrow R_j}$ is the response of $(E_i,R_j)$ to the filter $r^{(m')}$. Superposing thus only perturbs the targets by $(1-c)$ times the crosstalk term (with this perturbation vanishing for $c=1$). Indeed, with the targets of \cref{fig:experiment_2_pairs}, the measured crosstalk of the superposed filters was only about 0.3--0.6\,dB higher than that of the joint design [squares in \cref{fig:experiment_2_pairs}(b)]. In practice, filters optimized once for individual routings can thus be superposed for any combination of routings without re-optimizing, at a small cost in crosstalk. More generally, the weights $w_{ij}$ can encode any MIMO routing between emitters and receivers, of which the two experiments shown here are only examples.

\textit{Discussion.} Jointly optimizing the filters of an active RIS with an FFT-accelerated conjugate-gradient method, which scales to large systems, lowers crosstalk well below that of TR on the same experiment. Unlike TR, which only needs the Green's functions of the routed pairs, our design needs those of every receiver it protects from crosstalk.
Several directions remain open. One could formulate the objective differently, for instance with worst-case constraints on the crosstalk. For scenes with moving talkers, the short design time of our algorithm could allow the kernels to be optimized in real time, especially when using warm starts. Practical applications would also benefit from fewer Green's function measurements, for instance by interpolating between receiver positions or estimating the channels from ordinary speech. Finally, the proposed framework could be applied to other wave systems (such as microwave or radio communications).

\textit{Acknowledgments.} This work was supported in part by the Simons Foundation/Collaboration on Symmetry-Driven Extreme Wave Phenomena, by the U.S. Army Research Office through the Institute for Soldier Nanotechnologies (award no.~W911NF-23-2-0121), and by the Agence Nationale de la Recherche (award no.~ANR-22-CE24-0012).  R.P. acknowledges support from the U.S. National Science Foundation (award no.~IIS-2435905) and from the National Institute of Biomedical Imaging and Bioengineering of the National Institutes of Health (award no.~R21EB036343).

\textit{Data availability.} Code and measured data to reproduce our results are available at \url{https://github.com/FredericJorgensen/cocktail-party-ris-design}.

\bibliography{main}

\end{document}